\documentclass[aoas,MSNbibl,nameyear,dvips]{arximspdf}
\usepackage{multirow}
\usepackage{dcolumn}
\usepackage{graphicx}

\doi{10.1214/11-AOAS507}
\volume{6}
\issue{1}
\pubyear{2012}
\firstpage{27}
\lastpage{54}

\makeatletter

\setattribute{copyright}{owner}{Public Domain}

\newcolumntype{d}[1]{D{.}{.}{#1}}

\newcommand{\eqref}[1]{(\ref{#1})}

\newcommand{\var}{\operatorname{var}}
\newcommand{\cov}{\operatorname{cov}}

\newcommand{\epn}{\mathrm{E}}

\makeatother

\begin{document}
\begin{frontmatter}

\title{Marginal analysis of longitudinal count data in long sequences:
Methods and applications to a~driving study\thanksref{T1}}
\runtitle{Marginal analysis of longitudinal count data}

\thankstext{T1}{Supported by the Intramural Research Program of the National Institutes
of Health (NIH), Eunice Kennedy Shriver National Institute of Child
Health and Human Development. The computation was facilitated by the
Biowulf cluster computer system made available by the Center for
InformationTechnology at the NIH.}

\begin{aug}
\author[A]{\fnms{Zhiwei} \snm{Zhang}\corref{}\ead[label=e1]{zhiwei.zhang@fda.hhs.gov}},
\author[B]{\fnms{Paul S.} \snm{Albert}\ead[label=e2]{albertp@mail.nih.gov}}
\and
\author[C]{\fnms{Bruce} \snm{Simons-Morton}\ead[label=e3]{mortonb@mail.nih.gov}}
\runauthor{Z. Zhang, P. S. Albert and B. Simons-Morton}
\affiliation{Eunice Kennedy Shriver National Institute of
Child Health and~Human~Development}
\address[A]{Z. Zhang\\
Division of Biostatistics\\
Center for Devices\\
\quad and Radiological Health\\
Food and Drug Administration\\
10903 New Hampshire Ave.\\
Silver Spring, Maryland 20993\\
USA\\
\printead{e1}} 
\address[B]{P. S. Albert\\
Biostatistics and Bioinformatics Branch\\
Division of Epidemiology, Statistics\\
\quad and Prevention Research\\
Eunice Kennedy Shriver National Institute\\
\quad of Child Health and Human
Development\\
6100 Executive Blvd.\\
Bethesda, Maryland 20892\\
USA\\
\printead{e2}}
\address[C]{B. G. Simons-Morton\\
Prevention Research Branch\\
Division of Epidemiology, Statistics and Prevention Research\\
Eunice Kennedy Shriver National Institute\\
\quad of Child Health and Human
Development\\
6100 Executive Blvd.\\
Bethesda, Maryland 20892\\
USA\\
\printead{e3}}
\end{aug}

\received{\smonth{2} \syear{2011}}
\revised{\smonth{9} \syear{2011}}

%
\begin{abstract}
Most of the available methods for longitudinal data analysis are
designed and validated for the situation where the number of subjects
is large and the number of observations per subject is relatively
small. Motivated by the Naturalistic Teenage Driving Study (NTDS),
which represents the exact opposite situation, we examine standard and
propose new methodology for marginal analysis of longitudinal count
data in a small number of very long sequences. We consider standard
methods based on generalized estimating equations, under working
independence or an appropriate correlation structure, and find them
unsatisfactory for dealing with time-dependent covariates when the
counts are low. For this situation, we explore a within-cluster
resampling (WCR) approach that involves repeated analyses of random
subsamples with a final analysis that synthesizes results across
subsamples. This leads to a novel WCR method which operates on
separated blocks within subjects and which performs better than all of
the previously considered methods. The methods are applied to the NTDS
data and evaluated in simulation experiments mimicking the NTDS.
\end{abstract}

%

\end{frontmatter}

\section{Introduction}

\begin{figure}

\includegraphics{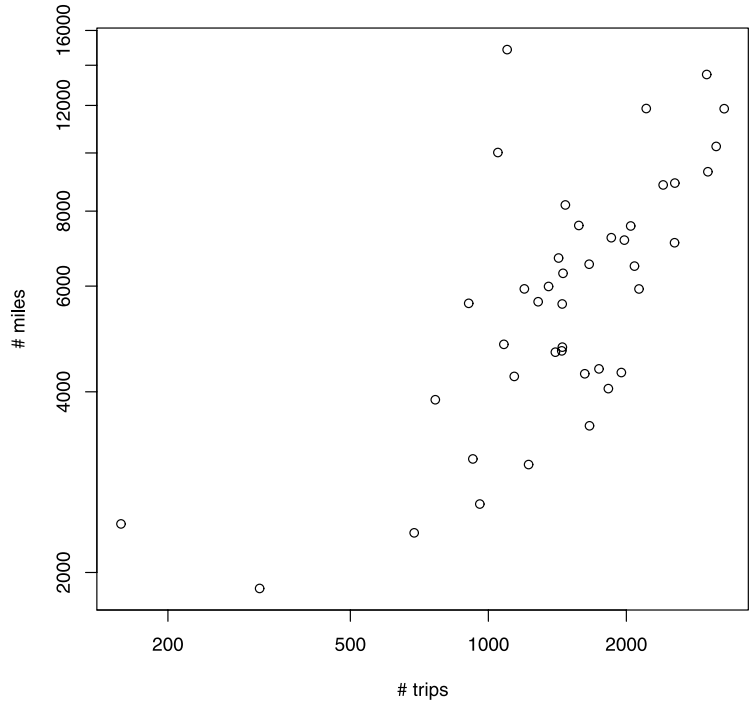}

\caption{Summary of NTDS data by subject: the number of
trips made and the total number of miles driven, both plotted on the
logarithmic scale.} \label{fig1}
\end{figure}

In this paper we consider the analysis of longitudinal data that arise
in the form of a small number of long sequences. Our interest in this
problem is motivated by the Naturalistic Teenage Driving Study (NTDS),
an observational study of teenage driving performance and
characteristics [Simons-Morton et al. (\citeyear{s11a,s11b})]. In the
study, 42 newly licensed
teenage drivers in Virginia were monitored continuously during their
first 18 months of independent driving using in-vehicle data recording
systems. The instrumentation included accelerometers, video cameras, a
global positioning system, a front radar and a lane tracker. The NTDS
is a first of its kind, at least for teenage drivers in the United
States. The study provides valuable information on risky driving
behavior, which can be assessed in terms of elevated gravitational
force ($g$-force) events (rapid start, hard stop, hard turn and yaw).
Counts of $g$-force events are available for each trip (defined as
ignition on to ignition off), and their incidence rates represent
different aspects of risky driving behavior. The NTDS data set
comprises more than 68,000 trips by the 42 teen subjects, with an
average of 1,626 trips per subject. Figure \ref{fig1} provides a summary
of the NTDS data by subject in terms of the number of trips made and
the total number of miles driven. An important goal in our analysis of
the NTDS data is to understand how risky driving is associated with
subject-level covariates (i.e., individual characteristics such as
gender) as well as trip-level or time-dependent covariates (e.g., time
since licensure, presence of passengers).

%

There is an extensive literature on longitudinal data analysis; see,
for example, \citet{d02}, \citet{f08} and \citet{m08}. Common
approaches to longitudinal data analysis include random effect models
[\citet{l82}, \citet{m08}] and generalized estimating
equations (GEE)
[\citet{l86}, \citet{z86}, \citet{zla88}]. In general,
the two approaches have different
interpretations (subject-specific versus marginal), although that
distinction is not important when the log link is used for count data.
For analyzing the NTDS data, it is natural to consider generalized
linear mixed models (GLMM) with appropriate random effects to account
for population heterogeneity, serial correlation and/or possible
overdispersion [\citet{c95}, \citet{m97}, \citet{m08}].
However, a realistic GLMM
for the
NTDS data would involve several random components, including a latent
process that induces serial correlation (see Section \ref{analysis}),
and the computational demand of such a GLMM analysis can be prohibitive
with thousands of trips per subject. Even if the computation is
feasible, the resulting inference may be sensitive to modeling
assumptions that are difficult to verify. The computational burden can
be reduced by resorting to a marginal analysis using the GEE approach,
especially under a working independence assumption. The GEE approach
only requires specification of the first two moments and not the entire
distribution. The robust variance estimate can be used to make
asymptotically valid inference that only requires correct specification
of the mean structure, assuming that the number of subjects is large
relative to the number of observations per subject. The exact opposite
situation occurs in the NTDS, and \citet{a95} have shown that the
robust variance estimate can perform poorly in such a situation and
that the model-based variance estimate may be preferable. Numerous
methods have been proposed to improve upon the robust variance
estimate, including jackknife [\citet{p88}, \citet{l90}],
bias correction
[\citet{m01}] and window subsampling techniques [\citet{s96},
\citet{h00}].

Given the well-known advantages and potential issues of the GEE
approach, it is natural to ask how to perform a simple and valid
marginal analysis of the NTDS data with reasonable efficiency and
robustness. We attempt to address this question in the present paper by
examining some existing methods and developing new ones. We find that
it is generally helpful to include a fixed effect for each subject in
the GEE model. Once the subject effects are estimated, they can be
treated as the response variable in a subsequent linear regression
analysis for estimating the effects of subject-level covariates. These
fixed effects also help with trip-level covariates by removing the
correlation due to population heterogeneity. If the counts are large
(with a marginal mean of 1, say), the effects of trip-level covariates
can be estimated from a conventional GEE analysis using an estimated
covariance matrix together with the robust variance estimate. For small
counts (with a marginal mean of 0.1), the conventional approach is not
satisfactory, and we explore a within-cluster resampling (WCR) approach
[\citet{h01}, \citet{f03}]. The WCR approach was originally
proposed to
deal with informative cluster sizes, which is not of concern in the
NTDS. Our motivation for considering WCR is to improve the performance
of GEE methods by altering the data structure. To this end, we consider
extensions of WCR that reduce serial correlation within clusters or
increase the number of (approximately) independent clusters. This leads
to a WCR method involving separated blocks which performs better than
the conventional methods.

The rest of the paper is organized as follows. In the next section we
set up the notation, formulate the problem, and discuss potential
issues. Then we examine the standard GEE methods in Section \ref{gee}
and explore some WCR methods in Section~\ref{mo}. The NTDS data are
analyzed in Section \ref{analysis} using the appropriate methods. The
paper concludes with a discussion in Section \ref{disc}.

\section{Formulation}\label{form}
Let $Y_{ij}$ ($i=1,\ldots,n$; $j=1,\ldots,k_i$) denote the number of
events that occur during the $j$th trip by the $i$th subject. For
reasons that will become clear later, we distinguish subject-level
covariates such as gender from trip-level covariates such as time since
licensure, writing $Z_i$ for the former group of covariates and
$X_{ij}$ for the latter. Note that $X_{ij}$ may include interactions
between subject-level and trip-level covariates. With $m_{ij}$ denoting
the mileage of the $j$th trip by the $i$th subject, the marginal model
of interest to us may be written as
%
\begin{equation}\label{10}
\epn(Y_{ij})=m_{ij}\exp(\nu+\alpha'Z_i+\beta'X_{ij}),
\end{equation}
where $\nu$, $\alpha$ and $\beta$ are unknown parameters, the latter
two being of primary importance. Without loss of generality, we treat
the covariates as fixed.

Estimation of the parameters in model \eqref{10} is challenged by the
fact that the $Y_{ij}$ from the same subject tend to be correlated due
to considerable variability between drivers as well as serial
correlation (over time) within drivers. These sources of correlation
are often accounted for using random effects [e.g., \citet
{z88}, \citet{d00}]. For example, one might postulate that,
conditional on the
random effects $b_i$, $c_{ij}$ and~$e_{ij}$, the $Y_{ij}$ are
independent and each follows a Poisson distribution with conditional mean
%
\begin{equation}\label{20}
\epn(Y_{ij}|b_i,c_{ij},e_{ij})=m_{ij}\exp(\nu^*+\alpha'Z_i+\beta
'X_{ij}+b_i+c_{ij}+e_{ij}),
\end{equation}
where $b_i$ induces some heterogeneity between drivers (beyond that
explained by~$Z_i$), $c_{ij}$ generates serial correlation among trips
close in time, and $e_{ij}$ accounts for any additional overdispersion
relative to the Poisson model. It is often assumed that the $c_{ij}$
for a given subject arise from a subject-specific stochastic process
$c_i$ through the relationship $c_{ij}=c_i(t_{ij})$, where $t_{ij}$
denotes the time of the $j$th trip. Note that model \eqref{20} is
consistent with model~\eqref{10} because it implies the same mean
structure after integrating out the random effects, provided the
distributions of the random effects do not depend on the covariates.

For the NTDS data, it seems reasonable to equip model \eqref{20} with
the following distributional assumptions. One might assume that
$b_i\sim N(0,\sigma_b^2)$, $e_{ij}\sim N(0,\sigma_e^2)$, and $c_i$ is a
zero-mean Gaussian process with a covariance structure given by
%
\begin{equation}\label{50}
\cov\{c_i(t_1),c_i(t_2)\}=\sigma_c^2\exp(-\gamma|t_1-t_2|).
\end{equation}
The random effects $b_i$ and $e_{ij}$ and the process $c_i$ are assumed
independent of each other, although the $c_{ij}=c_i(t_{ij})$ are
necessarily correlated within each subject. The parameter $\gamma>0$
determines how rapidly the serial correlation decreases with the gap
time. The process $c_i$ is known as the Ornstein--Uhlenbeck process
[\citet{u30}], and the above distributional assumptions will be
collectively referred to as the Gauss--Ornstein--Uhlenbeck--Poisson (GOUP)
model. It is possible to perform a maximum likelihood analysis under
the GOUP model [e.g., \citet{c95}]. However, such an analysis can
be computationally demanding because of the serial correlation, and the
resulting inference may be sensitive to the distributional assumptions
involved. The methods to be considered for our marginal analysis will
not require the full strength of the GOUP model, and they may or may
not involve the distributional assumptions in the GOUP model. Some of
the methods we consider do incorporate such distributional assumptions
into estimating equations through a working covariance matrix, in which
case we also assess the robustness of the resulting inference against
misspecification of the correlation structure. Our goal is to make
valid and reasonably efficient inference on $\alpha$ and $\beta$ in
model \eqref{10} with minimal dependence on the distributional
assumptions in the GOUP model.

Without the distributional assumptions, model \eqref{20} does play a
crucial role in this paper, as a way to conceptualize the different
sources of variability. We assume that the different subjects are
independent of each other and that the random effects $b_i$ and
$(e_{ij})_{j=1}^{k_i}$ and\vspace*{2pt} the random process $c_i$ are independent of
each other within each subject. It follows that the $Y_{ij}$ ($j=1,\ldots
,k_i$) are conditionally independent given $b_i$ and $c_i$. Another
important implication is that\looseness=-1
%
\begin{equation}\label{30}
\epn(Y_{ij}|b_i)=m_{ij}\exp(\nu_i+\beta'X_{ij}),
\end{equation}\looseness=0
where
%
\begin{equation}\label{40}
\nu_i=\nu^*+\alpha'Z_i+b_i+\log[\epn\{\exp(c_{ij}+e_{ij})\}]=\mbox
{const.}+\alpha'Z_i+b_i.
\end{equation}
This shows that the correlation due to $b_i$ can be removed by treating
subject as a fixed effect. Obviously, this approach would not work in
the usual asymptotic theory assuming a large number of subjects and a
limited number of observations per subject. However, it might be
appropriate when the number of subjects is small and the number of
observations per subject is large as in the NTDS. 
Note that the subject-specific model \eqref{30} does not directly
involve $\alpha$, which has been absorbed into the subject-specific
intercept~$\nu_i$; thus,~$\alpha$ cannot be estimated directly by
fitting model \eqref{30}. Nonetheless, equation~\eqref{40} suggests
that once $\nu_i$ has been estimated, say, by $\widehat\nu_i$, it
should be possible to estimate $\alpha$ from a subject-level linear
model regressing $\widehat\nu_i$ on~$Z_i$.

\section{Standard GEE methods}\label{gee}
\subsection{GEE with working independence}\label{wind}
Assuming working independence, a~standard GEE analysis can be performed
with or without fixed subject effects (FSE), using either the robust
variance estimate or the model-based variance estimate. This gives rise
to four possible methods for estimating~$\beta$. As discussed earlier,
$\alpha$ is not directly estimable from a GEE analysis with FSE but can
be recovered from a subsequent linear regression analysis based on
\eqref{40}. Let $\widehat\nu_i$ denote the GEE estimate of $\nu_i$, and
assume that $\widehat\nu_i$ is approximately unbiased, that is, $\epn
(\widehat\nu_i|\nu_i)\approx\nu_i$. This, together with \eqref{40},
implies that
\[
\epn(\widehat\nu_i)=\epn\{\epn(\widehat\nu_i|\nu_i)\}\approx\epn(\nu
_i)=\mbox{const.}+\alpha'Z_i.
\]
An approximately unbiased estimate of $\alpha$ can then be obtained
from a~sub\-ject-level linear model regressing $\widehat\nu_i$ on $Z_i$.
A standard least squares (LS) algorithm can be used to fit this model
if the $\widehat\nu_i$ can be treated as independent. In general, the
$\widehat\nu_i$ are correlated due to correlated errors in GEE
estimation, even though the $\nu_i$ are indeed independent of each
other. This correlation can be taken into account using an iteratively
reweighted least squares (IRLS) algorithm described in Appendix \ref{appA}.
Thus, under working independence, $\alpha$~can also be estimated using
four different methods (LS and IRLS with FSE, robust and model-based
without FSE).

These methods are evaluated and compared in a simulation study
mimicking the NTDS. Specifically, each simulated data set consists of
40 subjects and 60,000 trips (1500 per subject). The study duration is
rescaled to the unit interval, over which the trips are uniformly
distributed. The offset~$\log(m_{ij})$ follows a normal distribution
with mean~1 and variance~1,~$Z_i$ is generated as a Bernoulli variable
with success probability~0.5, and~$X_{ij}$ is taken to be the trip time
(i.e., $X_{ij}=t_{ij}$). Given the covariates, the outcome~$Y_{ij}$ is
generated according to the GOUP model described in Section~\ref{form},
with $\sigma_b^2=\sigma_c^2=\sigma_e^2=1$ and $\gamma=50$ or 300,
corresponding to longer- and shorter-lived serial correlation,
respectively. These values are based on the NTDS data (see Section \ref
{analysis}) and the wide range for $\gamma$ reflects a large amount of
variability in its estimation (see Section \ref{ecm}). We set $\alpha
=\beta=0$ for simplicity because the results change little over a range
of realistic values for these parameters. We choose $\nu^*$ in \eqref
{20} such that the marginal mean of~$Y_{ij}$ equals a~specified value
(0.1 or 1). This range for $\epn(Y_{ij})$ covers most kinematic
measures in the NTDS with varying thresholds. Some measures are
associated with larger counts, which are generally easier to deal with.
In each scenario (combination of parameter values), 1,000 replicate
samples are generated and analyzed using the methods described in the
preceding paragraph.

%
\begin{table}
\caption{Estimation of $\alpha$, the effect of a subject-level
covariate, using GEE methods with working independence. The methods are
described in Section \protect\ref{wind} and Appendix \protect\ref{appA}, and compared in
terms of empirical bias, standard deviation (SD), median standard error
(SE) and coverage probability (CP) of intended 95\% confidence
intervals. Each entry is based on 1,000 replicates}\label{tb1}
\begin{tabular*}{\tablewidth}{@{\extracolsep{\fill}}lcccd{2.2}ccc@{}}
\hline
\multicolumn{1}{@{}l}{\textbf{Mean}} & \multicolumn{1}{c}{\textbf{Serial}}
& \multicolumn{1}{c}{\textbf{Include}} & \multicolumn{1}{c}{\textbf{Further}}
& \multicolumn{1}{c}{\textbf{Emp.}} & \multicolumn{1}{c}{\textbf{Emp.}}
& \multicolumn{1}{c}{\textbf{Median}} & \multicolumn{1}{c@{}}{\textbf
{Emp.}}\\
\multicolumn{1}{@{}l}{\textbf{count}} & \multicolumn{1}{c}{\textbf{correlation}}
& \multicolumn{1}{c}{\textbf{FSE?}} & \multicolumn{1}{c}{\textbf{option}}
& \multicolumn{1}{c}{\textbf{bias}} & \multicolumn{1}{c}{\textbf{SD}}
& \multicolumn{1}{c}{\textbf{SE}} & \multicolumn{1}{c@{}}{\textbf{CP}}\\
\hline
1&Short&No&Robust&0.00&0.39&0.33&0.90\\
&&No&Model-based&0.00&0.40&0.03&0.14\\
&&Yes&LS&0.00&0.33&0.32&0.95\\
&&Yes&IRLS&0.01&0.32&0.31&0.95\\
&Long&No&Robust&0.00&0.42&0.34&0.90\\
&&No&Model-based&-0.01&0.42&0.03&0.11\\
&&Yes&LS&-0.01&0.33&0.33&0.94\\
&&Yes&IRLS&0.00&0.32&0.32&0.95\\[4pt]
0.1&Short&No&Robust&0.02&0.41&0.33&0.90\\
&&No&Model-based&0.01&0.40&0.04&0.17\\
&&Yes&LS&-0.01&0.35&0.33&0.94\\
&&Yes&IRLS&-0.01&0.36&0.32&0.93\\
&Long&No&Robust&-0.01&0.42&0.34&0.89\\
&&No&Model-based&0.00&0.40&0.04&0.15\\
&&Yes&LS&0.01&0.35&0.33&0.94\\
&&Yes&IRLS&0.00&0.33&0.32&0.95\\
\hline
\end{tabular*}
\end{table}

Table \ref{tb1} compares the four methods for estimating $\alpha$ in
terms of empirical bias, standard deviation, median standard error and
coverage probability of intended 95\% confidence intervals. All methods
in Table \ref{tb1} are nearly unbiased, suggesting that bias is not of
concern here. In terms of efficiency, the most important factor is
clearly the use of FSE, which consistently results in smaller standard
deviations. With FSE in the model, one might expect the IRLS estimate
of $\alpha$, which accounts for the correlation among the $\widehat\nu
_i$, to be more efficient than the LS estimate. However, their
difference seems rather small in Table \ref{tb1}, for two reasons. First, the
variability due to estimating the $\nu_i$ is dominated by the
variability in the $\nu_i$, resulting in weak correlation among the
$\widehat\nu_i$ in this particular example. Second, a referee pointed
out that any correlation that does exist among the $\widehat\nu_i$
should be approximately exchangeable, in which case the LS estimate is
still efficient. The precision in estimating $\alpha$ appears
insensitive to the length of the serial correlation (specified through
$\gamma$), although the FSE estimates do seem to become more variable
for low counts [$\epn(Y_{ij})=0.1$]. The use of FSE also helps with
variance estimation. Without FSE, the model-based variance estimate is
clearly disastrous, as expected, and even the robust variance estimate
is unsatisfactory, with sub-nominal coverage ($\approx$90\%). Including
FSE in the GEE model leads to reasonable variance estimates and nearly
correct coverage, under both (LS and IRLS) approaches. Thus, it seems
that the key to valid and efficient inference about $\alpha$ in similar
situations is to include FSE in a~GEE analysis followed by a
subject-level linear regression analysis for the estimated FSE.

%
\begin{table}[t!]
\caption{Estimation of $\beta$, the effect of a trip-level covariate,
using standard GEE methods. The methods are described in Sections
\protect\ref{wind}--\protect\ref{mcm} and compared in terms of
empirical bias, standard
deviation (SD), median standard error (SE), coverage probability (CP)
of intended 95\% confidence intervals and the percentage of estimates
that are not available (NA) due to numerical problems such as
noninvertible matrices. In each scenario, 1,000 replicates are
generated, and the NA estimates are counted and then excluded in
calculating the other summary statistics}\label{tb2}
\begin{tabular*}{\tablewidth}{@{\extracolsep{\fill}}lcccd{2.2}cccc@{}}
\hline
\multicolumn{1}{@{}l}{\textbf{Mean}} & \multicolumn{1}{c}{\textbf
{Serial}} & \multicolumn
{1}{c}{\textbf{Include}} & \multicolumn{1}{c}{\textbf{Variance}}&
\multicolumn{1}{c}{\textbf{Emp.}} & \multicolumn{1}{c}{\textbf{Emp.}}
& \multicolumn
{1}{c}{\textbf{Median}} & \multicolumn{1}{c}{\textbf{Emp.}}
&\\
\multicolumn{1}{@{}l}{\textbf{count}}
& \multicolumn{1}{c}{\textbf{correlation}} & \multicolumn
{1}{c}{\textbf{FSE?}} & \multicolumn{1}{c}{\textbf{estimate}}&
\multicolumn{1}{c}{\textbf{bias}} & \multicolumn{1}{c}{\textbf{SD}} &
\multicolumn
{1}{c}{\textbf{SE}} & \multicolumn{1}{c}{\textbf{CP}} & \multicolumn
{1}{c@{}}{\textbf{\%NA}}\\
\hline
\multicolumn{9}{@{}c@{}}{\textit{Working independence}}\\[3pt]
1&Short&No&Robust&0.00&0.12&0.09&0.91&0.0\\
&&No&Model-based&0.00&0.12&0.05&0.67&0.0\\
&&Yes&Robust&-0.01&0.12&0.09&0.90&0.0\\
&&Yes&Model-based&0.00&0.12&0.04&0.47&0.0\\
&Long&No&Robust&0.01&0.21&0.16&0.92&0.0\\
&&No&Model-based&-0.01&0.21&0.05&0.41&0.0\\
&&Yes&Robust&-0.01&0.21&0.16&0.91&0.0\\
&&Yes&Model-based&-0.01&0.21&0.04&0.28&0.0\\[3pt]
0.1&Short&No&Robust&0.00&0.13&0.10&0.90&0.0\\
&&No&Model-based&-0.01&0.13&0.07&0.74&0.0\\
&&Yes&Robust&0.00&0.13&0.10&0.90&0.0\\
&&Yes&Model-based&0.00&0.13&0.06&0.65&0.0\\
&Long&No&Robust&0.02&0.21&0.17&0.90&0.0\\
&&No&Model-based&0.00&0.21&0.07&0.49&0.0\\
&&Yes&Robust&-0.01&0.21&0.17&0.91&0.0\\
&&Yes&Model-based&-0.01&0.21&0.06&0.42&0.0\\
[3pt]
\multicolumn{9}{@{}c@{}}{\textit{Estimated covariance matrix}}\\[3pt]
1&Short&Yes&Robust&0.00&0.07&0.07&0.95&0.0\\
&&Yes&Model-based&-0.01&0.07&0.07&0.95&0.0\\
&Long&Yes&Robust&-0.01&0.12&0.12&0.94&0.0\\
&&Yes&Model-based&0.00&0.12&0.11&0.90&0.1\\[3pt]
0.1&Short&Yes&Robust&0.00&0.09&0.10&0.92&4.6\\
&&Yes&Model-based&0.00&0.10&0.10&0.95&4.9\\
&Long&Yes&Robust&0.00&0.15&0.16&0.90&3.5\\
&&Yes&Model-based&-0.01&0.15&0.12&0.87&3.9\\
[6pt]
\multicolumn{9}{@{}c@{}}{\textit{Misspecified covariance matrix}}\\[3pt]
1&Varying&Yes&Robust&-0.02&0.09&0.09&0.93&0.0\\
&&Yes&Model-based&-0.03&0.09&0.08&0.92&0.0\\[3pt]
0.1&Varying&Yes&Robust&-0.02&0.11&0.12&0.90&3.7\\
&&Yes&Model-based&-0.02&0.11&0.11&0.93&2.9\\
\hline
\end{tabular*}
\vspace*{12pt}
\end{table}

Estimation of $\beta$ is a different story, as shown in the top section
of Table~\ref{tb2}. As in the case of estimating $\alpha$, bias is not
a major issue in estimating~$\beta$. The precision in estimating $\beta
$ depends mostly on the length of the serial correlation, with better
precision for shorter-lived serial correlation, and seems insensitive
to other factors (e.g., FSE, mean count). This is clearly different
from the situation in Table \ref{tb1}, and an intuitive explanation is
the following. In general, estimating the effect of a subject-level
covariate is essentially comparing one group of subjects with another,
while estimating the effect of a trip-level covariate is essentially
comparing one group of trips with another group of trips by the same
subjects. With $X_{ij}=t_{ij}$, estimation of $\beta$ is basically a
comparison of later trips with earlier trips, and it seems natural that
serial correlation has a larger impact on this comparison than on a
comparison of boys with girls, say. The latter comparison, on the other
hand, is more likely to benefit from the use of FSE to separate the
relevant information (i.e., overall incidence rates of individual
drivers) from the noise (i.e., within-subject variation). In Table \ref{tb2},
none of the four methods is satisfactory in terms of coverage, though
the robust variance estimate performs better than the model-based one.
Several alternatives to the robust variance estimate have also been
explored with little success (see Section \ref{disc}).

\subsection{GEE based on an estimated covariance matrix}\label{ecm}

We now consider GEE methods incorporating the covariance matrix for the
$Y_{ij}$. Under the GOUP model described in Section \ref{form}, it is
straightforward to show, by appealing to well-known properties of
normal and Poisson distributions, that
%
\begin{eqnarray}
\epn(Y_{ij})&=&m_{ij}\exp\{\nu^*+\alpha'Z_i+\beta'X_{ij}+(\sigma
_b^2+\sigma_c^2+\sigma_e^2)/2\}=:\mu_{ij},\nonumber\\
\label{60}
\var(Y_{ij})&=&\mu_{ij}+\mu_{ij}^2\{\exp(\sigma_b^2+\sigma_c^2+\sigma
_e^2)-1\},\\
\label{70}\qquad
\cov(Y_{ij},Y_{ij'})&=&\mu_{ij}\mu_{ij'}[\exp\{\sigma_b^2+\sigma_c^2\exp
(-\gamma|t_{ij}-t_{ij'}|)\}-1],\nonumber\\[-8pt]\\[-8pt]
&&\eqntext{(j\not=j').}
\end{eqnarray}
These expressions provide the marginal covariance matrix relevant in
a~GEE analysis without FSE. With FSE in the model, we need to condition
on $b_i$ and the relevant formulas become
%
\begin{eqnarray}
\epn(Y_{ij}|b_i)&=&m_{ij}\exp(\nu_i+\beta'X_{ij})=:\mu_{ij|b_i},\nonumber
\\
\label{80}
\var(Y_{ij}|b_i)&=&\mu_{ij|b_i}+\mu_{ij|b_i}^2\{\exp(\sigma_c^2+\sigma
_e^2)-1\},\\
\label{90}
\cov(Y_{ij},Y_{ij'}|b_i)&=&\mu_{ij|b_i}\mu_{ij'|b_i}[\exp\{\sigma_c^2\exp
(-\gamma|t_{ij}-t_{ij'}|)\}-1],\nonumber\\[-8pt]\\[-8pt]
&&\eqntext{(j\not=j'),}
\end{eqnarray}
with $\nu_i$ defined in Section \ref{form}. Note that $\sigma_b^2$ is
not involved in the covariance matrix in a GEE analysis with FSE.
Unknown parameters in the covariance matrix ($\sigma_c^2$, $\sigma
_e^2$, $\gamma$ and possibly $\sigma_b^2$) can be estimated by applying
moment methods and nonlinear regression techniques to residuals from a
preliminary GEE analysis with working independence (see Appendix \ref{appB} for
details). This preliminary GEE analysis can be performed with or
without FSE, regardless of the primary GEE analysis for estimating
$\beta$. With FSE in the preliminary GEE analysis, the aforementioned
techniques do not provide an estimate of~$\sigma_b^2$. If desired, an
estimate of~$\sigma_b^2$ can be obtained from the IRLS estimation of~$\alpha$
or simply as the error variance in the LS analysis, ignoring
the fact that $\widehat\nu_i$ is an error-prone estimate of $\nu_i$
(see Section \ref{wind} and Appendix \ref{appA} for details). Thus, the
covariance matrix can be estimated using three methods: FSE-LS,
FSE-IRLS and no FSE, with the first two differing only in the estimate
of~$\sigma_b^2$.

%
\begin{table}
\tabcolsep=0pt
\caption{Estimation of parameters in the covariance matrix
$(\sigma_b^2,\sigma_c^2,\sigma_e^2,\gamma)$ using moment and nonlinear
regression methods. The methods are described in Appendix \protect\ref{appB} and
compared in terms of empirical bias, standard deviation (SD) and the
percentage of estimates that are not available (NA) due to convergence
failure in nonlinear regression. In each scenario, 1,000 replicates are
generated, and the NA estimates are counted and then excluded in
calculating the other summary statistics}\label{tb3}\vspace*{-3pt}
{\fontsize{8.6pt}{11pt}\selectfont{
\begin{tabular*}{\tablewidth}{@{\extracolsep{\fill
}}lcd{2.2}d{2.2}d{2.2}cccccd{2.1}@{}}
\hline
\multicolumn{1}{@{}l}{\multirow{2}{50pt}[-8pt]{\textbf{Serial correlation}}}
& & \multicolumn{4}{c}{\textbf{Emp. bias}}
& \multicolumn{4}{c}{\textbf{Emp. SD}} & \\[-4pt]
& & \multicolumn{4}{c}{\hspace*{-1.5pt}\hrulefill}
& \multicolumn{4}{c}{\hrulefill} &
\\
& \multicolumn{1}{c}{\textbf{Method}}&
\multicolumn{1}{c}{$\bolds{\sigma_b^2}$} & \multicolumn{1}{c}{$\bolds
{\sigma
_c^2}$} & \multicolumn{1}{c}{$\bolds{\sigma_e^2}$}&
\multicolumn{1}{c}{$\bolds{\gamma}$} &
\multicolumn{1}{c}{$\bolds{\sigma_b^2}$} & \multicolumn{1}{c}{$\bolds
{\sigma
_c^2}$} & \multicolumn{1}{c}{$\bolds{\sigma_e^2}$}&
\multicolumn{1}{c}{$\bolds{\gamma}$} & \multicolumn{1}{c@{}}{\textbf{\%
NA}}\\
\hline
\multicolumn{11}{@{}c@{}}{$\mbox{\textit{Mean count}}=10$}\\
[3pt]
Short&FSE-LS&-0.01&-0.03&-0.02&1.1E$+$01&0.23&0.10&0.11&1.4E$+$02&0.1\\
\quad($\gamma
=300$)&FSE-IRLS&-0.02&-0.03&-0.02&1.1E$+$01&0.23&0.10&0.11&1.4E$+$02&0.1\\
&No FSE&-0.26&0.01&-0.06&4.1E$+$02&0.25&0.29&0.22&5.9E$+$03&6.8\\
Long&FSE-LS&0.03&-0.10&-0.02&2.0E$+$01&0.24&0.07&0.08&4.9E$+$01&0.2\\
\quad($\gamma
=50$)&FSE-IRLS&0.02&-0.10&-0.02&2.0E$+$01&0.24&0.07&0.08&4.9E$+$01&0.2\\
&No FSE&-0.30&0.01&-0.06&1.8E$+$02&0.26&0.40&0.17&2.1E$+$03&5.8\\
[3pt]
\multicolumn{11}{@{}c@{}}{$\mbox{\textit{Mean count}}=1$}\\
[3pt]
Short&FSE-LS&-0.01&-0.03&0.00&1.2E$+$02&0.23&0.21&0.23&2.9E$+$03&0.3\\
\quad($\gamma
=300$)&FSE-IRLS&-0.02&-0.03&0.00&1.2E$+$02&0.23&0.21&0.23&2.9E$+$03&0.3\\
&No FSE&-0.29&0.01&-0.06&2.6E$+$02&0.23&0.24&0.21&4.5E$+$03&8.7\\
Long&FSE-LS&0.03&-0.10&-0.02&3.4E$+$01&0.25&0.10&0.21&8.5E$+$01&0.4\\
\quad($\gamma
=50$)&FSE-IRLS&0.01&-0.10&-0.02&3.4E$+$01&0.24&0.10&0.21&8.5E$+$01&0.4\\
&No FSE&-0.30&0.05&-0.08&2.1E$+$02&0.28&0.97&0.20&1.5E$+$03&17.5\\
[3pt]
\multicolumn{11}{@{}c@{}}{$\mbox{\textit{Mean count}}=0.1$}\\
[3pt]
Short&FSE-LS&0.28&-0.03&-0.13&6.3E$+$02&2.97&0.51&1.18&4.1E$+$03&7.9\\
\quad($\gamma
=300$)&FSE-IRLS&0.25&-0.03&-0.13&6.3E$+$02&2.97&0.51&1.18&4.1E$+$03&7.9\\
&No FSE&-0.28&0.08&-0.12&6.5E$+$02&0.35&0.39&0.39&3.6E$+$03&30.4\\
Long&FSE-LS&0.14&-0.04&-0.14&1.0E$+$03&1.65&0.51&1.17&1.4E$+$04&12.9\\
\quad($\gamma
=50$)&FSE-IRLS&0.08&-0.04&-0.14&1.0E$+$03&1.24&0.51&1.17&1.4E$+$04&12.9\\
&No FSE&-0.34&0.14&-0.14&2.2E$+$03&0.33&0.58&0.35&3.3E$+$04&37.3\\
\hline
\end{tabular*}
\vspace*{-3pt}
}}
\end{table}

Table \ref{tb3} shows the performance of the above three methods for
estimating parameters in the covariance structure, in terms of bias,
standard deviation and convergence. The methods are evaluated in the
same scenarios as the previous simulation experiments, with the
addition of a higher mean count~(10) to show a more complete picture.
For a mean count of 10, the FSE methods are nearly unbiased for the
variance components but biased for $\gamma$ in a way that results in
underestimation of the length of the serial correlation. The bias for
$\gamma$ is larger for longer-lived serial correlation. The no FSE
method is more biased for $\gamma$ (in the same direction) and even
visibly biased for $\sigma_b^2$. With decreasing counts, the bias
problem becomes more severe, especially for $\gamma$, to the extent
that estimation of the serial correlation is meaningless for a mean
count of $0.1$. The problem is compounded by a~large amount of
variability in estimating $\gamma$. Another issue with low counts is
convergence in the nonlinear regression. For a mean count of $0.1$, the
convergence failure rates are approximately 8\% and 13\% (with short-
and long-lived serial correlation, resp.) for the FSE methods
and even worse for the no FSE method. The FSE methods appear more
reasonable than the no FSE method, and we will use FSE-LS, which is
simpler than FSE-IRLS, in the subsequent simulations.

GEE methods based on covariance matrices estimated through FSE-LS are
evaluated in the same setting described in Section \ref{wind}. When
the nonlinear regression in FSE-LS fails to converge, the data-driven
initial values will be taken as the final estimates [see equation \eqref
{b10} in Appendix \ref{appB}]. With thousands of trips per subject, inverting a
covariance matrix is time-consuming, and iterating until convergence in
the usual Fisher scoring algorithm [\citet{f08}, Section 3.2.4]
would be prohibitive. We therefore settle for a one-step estimate of
the regression coefficients based on just one iteration in the Fisher
scoring algorithm. Specifically, we start with a GEE analysis with
working independence, use the residuals to estimate parameters in the
covariance matrix, and update the regression coefficients just once
using the estimated covariance matrix. Under this approach, it is
necessary to include FSE\vadjust{\goodbreak} in each GEE analysis; otherwise meaningful
simulation results cannot be produced. This is not surprising because
inversion of large covariance matrices can be difficult when the
correlation is strong \mbox{(without~FSE)} and poorly estimated. This is not a
serious limitation because the use of FSE leads to better efficiency
and coverage anyway, which we have observed in simulations where the
true covariance matrix is used in GEE (results not shown). Once FSE are
included, numerical problems become much less frequent (3--5\% for a
mean count of 0.1, $\le$1\% for higher counts).

We have found that incorporating the covariance matrix into the
estimating equation does little to improve the estimation of $\alpha$
in terms of efficiency and coverage. We therefore omit the results for
estimating $\alpha$ and henceforth focus on estimating $\beta$. The
results for the latter, reported in the middle section of Table \ref
{tb2}, show that better efficiency is attained using an estimated
covariance matrix than under working independence. In terms of
coverage, the robust variance estimate performs well for higher counts
but not for lower ones, especially when the serial correlation is
long-lived. The model-based variance estimate performs well for
short-lived serial correlation and poorly for long-lived serial
correlation. The former case shows that underestimating the length of
the serial correlation has little effect when the serial correlation is
already negligibly short. Although not shown in Table \ref{tb2}, the
model-based variance estimate has been found to work well in all cases
if the true covariance matrix is used. Thus, the root of the problem
appears to be the difficulty in estimating $\gamma$.

\subsection{GEE based on a misspecified covariance matrix}\label{mcm}
It is natural to ask how the GEE methods based on the GOUP model would
perform if the model is misspecified. For the NTDS, the GOUP model
seems plausible for the most part. It is possible, however, that the
length of the serial correlation, which is assumed constant, might
change over time. To formalize this notion of changing length, note
that the covariance structure given by \eqref{50} can be generalized into
%
\begin{equation}\label{100}
\cov\{c_i(t_1),c_i(t_2)\}=\sigma_c^2\exp\biggl(-\int_{t_1}^{t_2}\gamma
(t)\,dt\biggr),\qquad t_1<t_2,
\end{equation}
for a function $\gamma\dvtx[0,1]\to(0,\infty)$. For a constant $\gamma$,
this reduces to the original GOUP model. In the next set of simulation
experiments, we will generate data using the generalized GOUP model
with $\gamma$ taken to be a straight line from $\gamma(0)=300$ to
$\gamma(1)=50$. Thus, instead of working exclusively with short- or
long-lived serial correlation, we allow it to be short-lived at the
beginning and to gradually become long-lived at the end. This is
motivated by the conjecture that teenage driving will tend to be
haphazard at the beginning and will become more consistent over time.
Unfortunately, it appears difficult to verify this conjecture directly
using the data. Indeed, even when~$\gamma$ is a constant, we have seen
in Table \ref{tb3} that it can be really difficult to estimate,\vadjust{\goodbreak}
especially when the counts are low. It seems reasonable to expect that
a time-varying $\gamma$ will be even more difficult to estimate. We
therefore take a sensitivity analysis approach to this issue. The
bottom section of Table \ref{tb2} shows the performance in the present
setting of GEE methods based on the original GOUP model. The point
estimates are more (or less) efficient in this setting than if the
serial correlation is consistently long-lived (or short-lived). The
coverage property of the model-based variance estimate is also
intermediate between the two extremes. As before, the robust variance
estimate works well for large counts but not for small ones.

\subsection{Summary}
To summarize, valid inference on $\alpha$ can be made through a GEE
analysis with FSE (regardless of the covariance matrix) followed by
a~linear regression analysis, while inference on $\beta$ requires more
work. For large counts, valid inference on $\beta$ can be made by
estimating the covariance matrix and using the robust variance
estimate; in this case, the inference seems robust with respect to the
serial correlation structure. For small counts, the robust variance
estimate is unsatisfactory and the model-based variance estimate
performs well only for short-lived serial correlation. Because the
serial correlation is difficult to estimate with small counts, one
would not know in practice whether the model-based variance estimate is
appropriate for a particular application. Therefore, it is necessary to
develop methods that perform well (or better) for small counts with
short- and long-lived serial correlation. This is the focus of the next
section, where we explore WCR-type methods.

\section{WCR and extensions}\label{mo}
\subsection{The standard WCR approach}\label{smo}
WCR is a resampling approach for marginal analysis of clustered data,
originally designed to deal with informative cluster sizes [\citet
{h01}, \citet{f03}]. In its original form, the approach can be
applied to the
present problem as follows. Suppose we sample a trip from each subject
randomly and form a subset of nonclustered data
%
\begin{equation}\label{110}
\{(Z_{iJ_i},X_{iJ_i},Y_{iJ_i}),i=1,\ldots,n\},
\end{equation}
where $J_i$ is uniformly distributed on $\{1,\ldots,k_i\}$,
independently across $i$. There is no correlation in the subsample
\eqref{110}, which can therefore be analyzed using a quasi-Poisson
model with mean structure given by \eqref{10}. This\vspace*{1pt} can be repeated a
large number (say, $L$) of times to yield $\{(\widehat\theta_l,\widehat
{\mathbf\Sigma}_l),l=1,\ldots,L\}$, where $\widehat\theta_l$ denotes the
point estimate of $\theta=(\alpha,\beta)$ based on the $l$th subsample,
and $\widehat{\mathbf\Sigma}_l$ is the associated variance
estimate.
Then the WCR estimate of~$\theta$ is given by
%
\begin{equation}\label{115}
\widehat\theta_{\mathrm{WCR}}=\frac1L\sum_{l=1}^L\widehat\theta_l,\vadjust{\goodbreak}
\end{equation}
and the associated variance estimate is
%
\begin{equation}\label{120}
\widehat{\mathbf\Sigma}_{\mathrm{WCR}}=\frac1L\sum_{l=1}^L\widehat
{\mathbf\Sigma}_l
-\frac1{L-1}\sum_{l=1}^L(\widehat\theta_l-\widehat\theta_{\mathrm
{WCR}})(\widehat\theta_l-\widehat\theta_{\mathrm{WCR}})'.
\end{equation}
The first term on the right-hand side of \eqref{120} measures the
variability within subsamples, while the second term measures the
variability between subsamples. 


In general, for any WCR method to work, it is essential that valid
estima\-tes $(\widehat\theta_l,\widehat{\mathbf\Sigma}_l)$ be obtained
from each subsample. Note that the estimates~$(\widehat\theta_l,\widehat
{\mathbf\Sigma}_l)$ are identically distributed. If $\widehat\theta_l$
is biased for $\theta$, then $\widehat\theta_{\mathrm{WCR}}$ given by~\eqref{115} is also biased by the same amount. Similarly, if $\widehat
{\mathbf\Sigma}_l$ underestimates the sampling variance of~$\widehat
\theta_l$, then $\widehat{\mathbf\Sigma}_{\mathrm{WCR}}$ given by \eqref
{120} will underestimate the sampling variance of $\widehat\theta
_{\mathrm{WCR}}$ by a similar amount, because the second term on the
right-hand side of \eqref{120} will be approximately unbiased for large
$L$. Furthermore, in terms of relative bias, the underestimation
problem of $\widehat{\mathbf\Sigma}_{\mathrm{WCR}}$ is generally worse
than that of $\widehat{\mathbf\Sigma}_l$, simply because the true
variance of $\widehat\theta_{\mathrm{WCR}}$ is smaller than that of
$\widehat\theta_l$. In light of these observations, we will study WCR
methods in two steps, always considering a single subsample before
moving to repeated sampling.

\subsection{WCR with simple random sampling (WCR-SRS)}\label{mo-srs}
There is no reason why a WCR subsample must consist of exactly one
observation from each subject. In fact, to fit a model with FSE, there
has to be more than one observation from each subject. Without FSE, the
standard WCR approach is feasible but, as we shall see, does not work
well. We therefore consider an extended WCR approach where a simple
random sample (SRS) is taken from each subject to form an WCR
subsample. Thus, instead of a single number $J_i$, we take an SRS $\{
J_i^{(1)},\ldots,J_i^{(R)}\}$ from the index set $\{1,\ldots,k_i\}$ for
each $i$. The resulting subsample is
\[
\bigl\{\bigl(Z_{iJ_i^{(r)}},X_{iJ_i^{(r)}},Y_{iJ_i^{(r)}}\bigr),i=1,\ldots,n,r=1,\ldots
,R\bigr\},
\]
for which a standard GEE analysis can be performed, under working
independence or using an estimated covariance matrix, with or without
FSE, using the robust or model-based variance estimate. This can be
repeated a~large number of times, and the resulting estimates can be
combined using~\eqref{115} and~\eqref{120}. If the covariance
matrix for the $Y_{ij}$ is used, it will be estimated once from the
full sample and the estimate will be applied to all subsamples.

%
\begin{table}
\tabcolsep=0pt
\caption{Estimation of $\beta$, the effect of a trip-level covariate,
using the WCR-SRS approach with different values of $R$ (number of
trips to sample from each subject) and $L$ (number of repetitions in
WCR). The methods are described in Sections \protect\ref{smo} and
\protect\ref{mo-srs}
and compared in terms of empirical bias, standard deviation (SD),
median standard error (SE), coverage probability (CP) of intended 95\%
confidence intervals and the percentage of estimates that are not
available (NA). The mean count is fixed at 0.1. In each scenario, 1,000
replicates are generated, and the NA estimates are counted and then
excluded in calculating the other summary statistics}\label{tb4}
\begin{tabular*}{\tablewidth}{@{\extracolsep{\fill
}}lccd{3.0}d{3.0}d{3.2}d{4.2}ccc@{}}
\hline
\multicolumn{1}{@{}l}{\textbf{Serial}} & \multicolumn{1}{c}{\textbf
{Include}} & \multicolumn
{1}{c}{\textbf{Variance}} & \multicolumn{2}{c}{}&
\multicolumn{1}{c}{\textbf{Emp.}} & \multicolumn{1}{c}{\textbf{Emp.}} &
\multicolumn
{1}{c}{\textbf{Median}} & \multicolumn{1}{c}{\textbf{Emp.}}
&\\
\multicolumn{1}{@{}l}{\textbf{correlation}} & \multicolumn{1}{c}{\textbf
{FSE?}} & \multicolumn
{1}{c}{\textbf{estimate}} & \multicolumn{1}{c}{$\bolds{R}$}&
\multicolumn{1}{c}{$\bolds{L}$} & \multicolumn{1}{c}{\textbf{bias}}
& \multicolumn{1}{c}{\textbf{SD}} & \multicolumn{1}{c}{\textbf{SE}}&
\multicolumn{1}{c}{\textbf{CP}} & \multicolumn{1}{c}{\textbf{\%NA}}\\
\hline
\multicolumn{10}{@{}c@{}}{\textit{Working independence}}\\[4pt]
Short&No&Robust&1&1&-15.40&357.22&1.59&0.64&0.4\\
&&&5&1&-0.02&1.44&0.99&0.86&0.0\\
&&&25&1&0.03&0.64&0.52&0.91&0.0\\
&&&100&1&-0.02&0.37&0.28&0.91&0.0\\
&&&500&1&-0.01&0.20&0.15&0.92&0.0\\
&&&100&500&0.00&0.13&0.00&0.16&0.0\\
&Yes&Robust&5&1&0.09&1.59&1.16&0.87&0.2\\
&&&25&1&0.01&0.68&0.50&0.87&0.1\\
&&&100&1&-0.02&0.37&0.28&0.90&0.1\\
&&&500&1&0.00&0.18&0.14&0.91&0.0\\
&&&100&500&0.00&0.12&0.00&0.18&0.0\\[4pt]
Long&No&Robust&1&1&32.12&2011.24&1.64&0.64&0.8\\
&&&5&1&-0.08&1.43&1.00&0.89&0.0\\
&&&25&1&-0.01&0.72&0.53&0.89&0.0\\
&&&100&1&0.01&0.42&0.31&0.92&0.0\\
&&&500&1&0.00&0.24&0.20&0.91&0.0\\
&&&100&500&-0.01&0.21&0.10&0.52&0.0\\
&Yes&Robust&5&1&-0.62&19.13&1.13&0.85&0.1\\
&&&25&1&-0.02&0.71&0.52&0.89&0.1\\
&&&100&1&0.00&0.39&0.30&0.90&0.0\\
&&&500&1&-0.01&0.24&0.20&0.90&0.0\\
&&&100&500&-0.01&0.21&0.10&0.56&0.0\\
\hline
\end{tabular*}
\end{table}

\setcounter{table}{3}
%
\begin{table}
\tabcolsep=0pt
\caption{(Continued)}
\begin{tabular*}{\tablewidth}{@{\extracolsep{\fill
}}lccd{3.0}d{3.0}d{2.2}cccc@{}}
\hline
\multicolumn{1}{@{}l}{\textbf{Serial}} & \multicolumn{1}{c}{\textbf
{Include}} & \multicolumn
{1}{c}{\textbf{Variance}} & \multicolumn{2}{c}{}&
\multicolumn{1}{c}{\textbf{Emp.}} & \multicolumn{1}{c}{\textbf{Emp.}} &
\multicolumn
{1}{c}{\textbf{Median}} & \multicolumn{1}{c}{\textbf{Emp.}}
&\\
\multicolumn{1}{@{}l}{\textbf{correlation}} & \multicolumn{1}{c}{\textbf
{FSE?}} & \multicolumn
{1}{c}{\textbf{estimate}} & \multicolumn{1}{c}{$\bolds{R}$}&
\multicolumn{1}{c}{$\bolds{L}$} & \multicolumn{1}{c}{\textbf{bias}}
& \multicolumn{1}{c}{\textbf{SD}} & \multicolumn{1}{c}{\textbf{SE}}&
\multicolumn{1}{c}{\textbf{CP}} & \multicolumn{1}{c}{\textbf{\%NA}}\\
\hline
\multicolumn{10}{@{}c@{}}{\textit{Estimated covariance matrix}}\\[4pt]
Short&Yes&Robust&5&1&0.03&1.44&0.89&0.75&0.4\\
&&&25&1&0.02&0.59&0.48&0.85&1.4\\
&&&100&1&0.01&0.30&0.29&0.91&3.3\\
&&&500&1&-0.01&0.15&0.15&0.92&4.6\\
&&&100&500&0.00&0.09&0.21&0.70&2.6\\
&&Model-based&5&1&-0.09&1.49&1.19&0.91&0.2\\
&&&25&1&0.02&0.58&0.55&0.94&1.1\\
&&&100&1&0.00&0.29&0.30&0.94&2.2\\
&&&500&1&0.00&0.15&0.15&0.97&4.1\\
&&&100&500&0.01&0.10&0.12&0.68&1.4\\[4pt]
Long&Yes&Robust&5&1&-0.04&1.48&0.93&0.76&0.4\\
&&&25&1&0.00&0.61&0.51&0.87&1.2\\
&&&100&1&0.00&0.31&0.30&0.90&2.3\\
&&&500&1&-0.01&0.18&0.19&0.90&3.1\\
&&&100&500&0.00&0.15&0.24&0.69&2.8\\
&&Model-based&5&1&-0.02&1.67&1.22&0.92&0.4\\
&&&25&1&0.01&0.59&0.54&0.93&1.9\\
&&&100&1&-0.01&0.32&0.31&0.94&2.4\\
&&&500&1&0.00&0.18&0.17&0.92&3.2\\
&&&100&500&0.00&0.16&0.16&0.72&2.6\\
\hline
\end{tabular*}
\end{table}

This WCR-SRS approach is evaluated in Table \ref{tb4}. Given the
findings in the last section, this evaluation is focused on small
counts. Under working independence, only the robust variance estimate
is considered because the model-based variance estimate is clearly
invalid. As before, no meaningful results are available when an
estimated covariance matrix is used without FSE. For each method, we
begin by analyzing a single subsample, with $L=1$ and different values
of $R$ (size of the SRS from each subject). Table \ref{tb4} shows poor
performance for $R=1$ (i.e., standard WCR), apparently due to erratic
estimates, which are likely to arise when the subsample is very small.
Even for $R=5$, the point estimate is visibly biased in some cases,
probably because the subsample is still small. The coverage probability
generally increases with $R$, at least for $R\le100$. The best coverage
is seen when an estimated covariance matrix is used together with FSE
and the model-based variance estimate. As explained in Section \ref
{ecm}, this method tends to work well when the serial correlation is
weak, which is precisely what happens to a randomly selected subset of
trips, which tend to be farther apart from each other than in the full
sample. This explains why better coverage is seen here than in the
corresponding portion of Table \ref{tb2}. Of course, this trend will be
reversed when the number of selected trips gets too large, which
explains the slight drop in coverage probability when $R$ increases
further to 500. Overall, it appears that $R=100$ would be a reasonable
choice in terms of bias and coverage. Efficiency is not a major
consideration here because the efficiency loss due to $R\le k_i$ can be
recovered through repeated sampling. Table~\ref{tb4} also shows the
performance of the WCR-SRS method with $L=500$ and $R=100$, whose
efficiency levels appear similar to the corresponding results in
Table~\ref{tb2}, as expected. For the working independence methods, the WCR
variance estimate is seriously biased downward and the coverage
probability is far below the nominal level, confirming the conjecture
in Section \ref{smo} that the variance underestimation problem of~%
$\widehat{\mathbf\Sigma}_{\mathrm{WCR}}$ is generally worse than that
of $\widehat{\mathbf\Sigma}_l$ in terms of relative bias. For methods
based on an estimated covariance matrix, the WCR variance estimate is
not necessarily biased downward but the coverage properties are not
good. This behavior is probably due to erratic estimates which arise
frequently in repeated analyses of subsamples when a poorly estimated
covariance matrix is used.

Table \ref{tb4} does suggest a potential benefit of WCR, that is, the ability
to produce a sparse subset of trips with weaker serial correlation.
This advantage can be exploited further by selecting trips that are
separated from each other by a specified amount. Specifically, for a
given positive integer $S$, we can choose a trip randomly among the
first $S+1$ trips and every $(S+1)$st trip from there on, until the end
of the sequence. This can be done for each subject and the process can
be repeated many times as before. This approach does lead to good
coverage in single outputation when an estimated covariance matrix is
used together with FSE and the model-based variance estimate (results
not shown). However, upon moving to WCR, this approach exhibits the
same strange behavior as seen in Table \ref{tb4} (results not shown),
probably for the same reason.

\subsection{WCR with separated blocks (WCR-SB)}\label{mosb}
The previous experiments with WCR show that variance underestimation
can be a serious problem, especially when the variance of $\widehat
\theta_{\mathrm{WCR}}$ is much smaller than that of $\widehat\theta_l$.
The latter difference can be reduced by including more observations in
each subsample, and the question then becomes how to ensure reasonable
performance of the variance estimate based on a subsample. To this end,
we propose the following WCR approach involving separated blocks. With
FSE in the model, the only source of correlation is serial correlation,
which is weaker for trips farther apart. Thus, with an appropriate
amount of separation, trips from the same subject can be treated as
approximately independent. Let $B$ denote the block size (i.e., the
number of consecutive trips to be analyzed together as a block) and $S$
the separation (i.e., the number of trips used to separate the blocks).
For a given subject, one possible set of separated blocks can be formed
by taking the first $B$ trips, skipping the next $S$ trips, taking the
next~$B$, skipping the next $S$, and so on. Each group of $B$ trips
sampled together will be treated as a block and the different blocks
will be treated as independent in the subsequent GEE analysis. A random
shift can be added to this sampling process, and the random sampling
process can be repeated many times as before.

Preliminary simulation results, as well as those in Table \ref{tb4},
suggest that a reasonable block size would be $B=100$. Simple
calculations based on \eqref{80} and \eqref{90} show that,
with $S=50$, the correlation between trips in different blocks is
typically well below 0.05 in all scenarios considered in this paper.
With these choices for $B$ and~$S$, the WCR-SB approach is evaluated
through simulations and the results are presented in Table \ref{tb5}.
%
%
\begin{table}
\caption{Estimation of $\beta$, the effect of a trip-level covariate,
using the WCR-SB approach with working independence and the robust
variance estimate, implemented with $B=100$ (block size), $S=50$
(separation) and $L=1$ (single subsample) or 50. The method is
evaluated in terms of empirical bias, standard deviation (SD), median
standard error (SE), coverage probability (CP) of intended 95\%
confidence intervals and the percentage of estimates that are not
available (NA). The mean count is fixed at 0.1. In each scenario, 1,000
replicates are generated, and the NA estimates are counted and then
excluded in calculating the other summary statistics}\label{tb5}
\begin{tabular*}{\tablewidth}{@{\extracolsep{\fill}}ld{2.0}ccccc@{}}
\hline
\multicolumn{1}{@{}l}{\textbf{Serial}} & &
\multicolumn{1}{c}{\textbf{Emp.}} & \multicolumn{1}{c}{\textbf{Emp.}} &
\multicolumn
{1}{c}{\textbf{Median}} & \multicolumn{1}{c}{\textbf{Emp.}}
&\\
\multicolumn{1}{@{}l}{\textbf{correlation}} & \multicolumn{1}{c}{$\bolds{L}$}&
\multicolumn{1}{c}{\textbf{bias}} & \multicolumn{1}{c}{\textbf{SD}} &
\multicolumn
{1}{c}{\textbf{SE}} & \multicolumn{1}{c}{\textbf{CP}} & \multicolumn
{1}{c@{}}{\textbf{\%NA}}\\
\hline
Short&1&0.00&0.15&0.13&0.94&0.0\\
&50&0.00&0.12&0.10&0.92&0.0\\
Long&1&0.00&0.22&0.19&0.92&0.0\\
&50&0.01&0.20&0.17&0.92&0.0\\
Varying&1&0.00&0.18&0.15&0.92&0.1\\
&50&0.00&0.15&0.13&0.92&0.0\\
\hline
\end{tabular*}
\end{table}
We focus on working independence to avoid numerical stability issues,
and only consider the robust variance estimate. The method appears to
have better coverage when applied to a single set of separated blocks
than to the full sample (Table \ref{tb2}), apparently owing to a larger
number of approximately independent clusters (400 vs. 40). Note that the
efficiency based on a single outputation is quite close to that in a
full sample analysis. Considering this and the computational demand,
the WCR version is implemented with $L=50$ repetitions, and the
resulting WCR-SB estimates are virtually as efficient as the full
sample estimates. In the case of long-lived serial correlation, better
coverage (92\%) is achieved here using the working independence method
than in any of the previous simulations. The improvement (2\% over
Table~\ref{tb2}) is not dramatic but still substantial, and similar
phenomena have been observed consistently in other simulation
experiments under similar scenarios. Table \ref{tb5} also shows the
performance of the WCR-SB method when the serial correlation is
generated according to \eqref{100} rather than \eqref{50}. The only
noticeable effect of this change is an intermediate level of efficiency
(between the extreme cases of short- and long-lived serial correlation).

\section{Analysis of NTDS data}\label{analysis}
Our analysis of the NTDS data concerns the following elevated $g$-force
events: rapid start (longitudinal $\mbox{acceleration}\ge0.35g$), hard stop
(longitudinal $\mbox{deceleration}\ge0.40g$), hard left/right turn (lateral
decel\-eration/$\mbox{acceleration} \ge0.45g$), yaw ($\ge$5 degrees within 3
seconds), as well as a composite measure defined as the totality of
these 5 types of events. Yaw is a measure of correction after a turn
and is calculated as the absolute change in angle between an initial
turn and the correction. For each specific $g$-force event, the threshold
(in parenthesis) is chosen to allow meaningful differences between
drivers and different driving conditions to be revealed and estimated
with a sufficient number of events. The mean counts based on the chosen
thresholds are 0.13 (rapid start), 0.18 (hard stop), 0.20 (hard left
turn), 0.15 (hard right turn), 0.04 (yaw) and 0.70 (composite measure).
The mean count for yaw is relatively low, perhaps too low for the
methods discussed in this paper, but the corresponding threshold is
already the lowest for which we have data available. This limitation
should be kept in mind when interpreting the results.

\begin{figure}[b]

\includegraphics{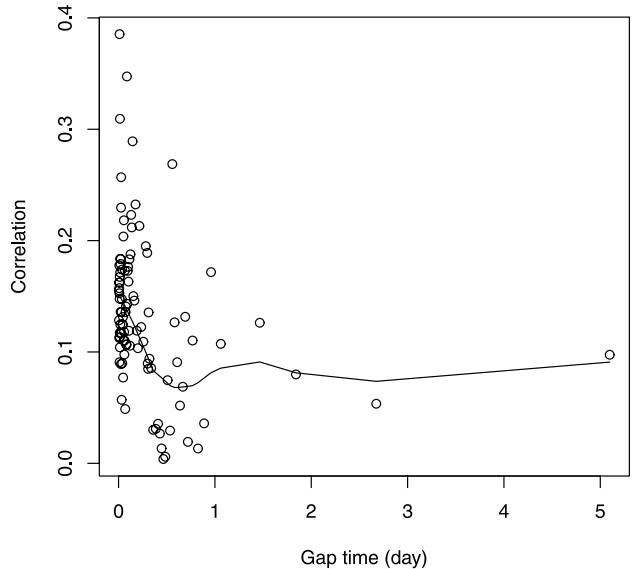}

\caption{Serial correlation for the composite measure in the NTDS.
Pairwise products of standardized residuals are calculated for
consecutive trips and, after sorting by gap time, grouped into 100
bins. Each circle represents the median gap time and the mean product
of standardized residuals in a bin, and the solid line is a lowess
smooth.} \label{fig2}
\end{figure}

From the viewpoint of behavior science, it is reasonable to expect some
serial correlation because risky driving may be related to the weather
and the driver's mental and physical conditions (e.g., mood and
fatigue). All of these factors could result in serial correlation
unless they are all included in the model, which is impossible in the
NTDS due to the lack of this information. To explore the nature of the
serial correlation in the NTDS data, a~marginal model with FSE and all
relevant covariates (to be described later) is fit for the composite
measure under working independence. Pairwise products of standardized
residuals are calculated for consecutive trips (because the number of
all possible pairs is too large). After sorting by gap time, these
pairs are grouped into 100 bins for further reduction. Within each bin,
we calculate the median gap time and the mean product of standardized
residuals, and the results are plotted in Figure \ref{fig2} together
%
with a~lowess smooth. Figure\vadjust{\goodbreak} \ref{fig2} suggests that some serial
correlation is present in the data, although a precise characterization
of the serial correlation seems difficult. Under the GOUP model and
using the techniques described in Appendix \ref{appB}, the estimated variance
components are generally close to 1 (for the composite measure and
other $g$-force events), while the estimate of $\gamma$ varies wildly and
depends heavily on the initial value.

%
\begin{figure}

\includegraphics{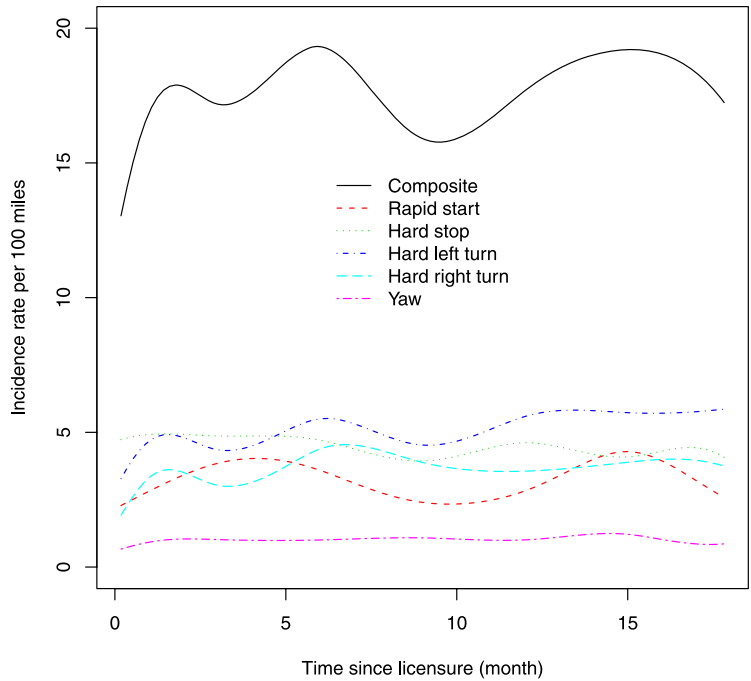}

\caption{Incidence rates of various $g$-force events (rapid start, hard
stop, hard left/right turn, yaw, and the composite measure) as
functions of time since licensure, estimated using regression splines
(one node for every 3 months).} \label{fig3}
\end{figure}

Figure \ref{fig3} presents the estimated incidence rate (IR) per 100
miles of each $g$-force event as a function of time since licensure. The
IR is defined by~\eqref{10} with $m_{ij}$ replaced by 100, $Z_i$ empty,
and $X_{ij}$ consisting of indicators of calendar month and a set of
regression splines for $t_{ij}$ (one knot for every 3 months). Calendar
month is included in the model to adjust for a possible seasonal
effect, which may confound with the effect of time since licensure
because the enrollment of subjects was not uniform with respect to
calendar month, with more subjects enrolled in the fall and fewer in
the spring. The estimates shown in Figure \ref{fig3} represent
geometric means of the monthly estimates. Note that the IR considered
here is a marginal quantity which involves the marginal intercept $\nu$
and which should therefore be estimated in a model without FSE. The
estimates in Figure \ref{fig3} are obtained under working independence,
because the GOUP model without FSE is problematic to fit and its
performance is not well understood (see Section \ref{ecm}). Under
the regression spline model, there is some ambiguity as to whether the
IR for the composite measure changes over time. The $p$-value for the
regression splines is 0.49 based on the robust variance estimate under
working independence, although the validity of this test may be
questionable (see Section \ref{gee}). Under the GOUP model with FSE,
the IR for the composite measure does appear to change over time
($p=0.0014$ for the robust variance estimate, $p<0.0001$ for the
model-based variance estimate). The latter observation is in contrast
to a previous report that risky driving largely remains constant
[\citet{s11a}], and the difference is probably due to the lower
thresholds used here, which lead to more events and perhaps more
relevant information. Despite the ambiguity about statistical
significance, the main conclusion from our marginal analysis is fairly
consistent with the findings of \citet{s11a}. The latter reference
reports IRs for adults/parents that are consistently and substantially
lower than those for novice teenagers driving the same vehicles. In
Figure \ref{fig3}, the estimated IR of the composite measure for
teenage drivers increases over the first 6 months, declines over the
next few months, and then increases again, suggesting the establishment
of a risky driving style. This general impression is reinforced by the
plots for the specific $g$-force events, which show some temporal changes
but no clear trends.

As noted by a referee, some instances of risky driving (e.g., mistakes
due to the lack of skills) may depend more on the amount of practice,
which can be quantified by accumulated mileage, than on time since
licensure. This raises the question of which measure is more
appropriate to adjust for. Accumulated mileage is a measure of driving
experience, while time since licensure reflects both experience and
maturity. From a public health perspective, we believe that time since
licensure is more relevant to policy makers. Nonetheless, all analyses
in this section have been repeated with time since licensure replaced
by accumulated mileage, and the results are very similar and therefore
omitted. With regard to Figure \ref{fig3}, this similarity suggests
that persistent risky driving among teenage drivers is not only due to
inexperience but also to immaturity (i.e., aspects of adolescent
development that motivate novice young drivers to seek excitement and
under-recognize risks).

Each $g$-force event is further analyzed in a larger model that
simultaneously adjusts for the following covariates: driver's gender,
risky friends, time of day, passenger condition, calendar month, and
time since licensure. The risky behavior of a teenage driver's friends
is assessed using a 7-item index asking ``How many of your friends would
you estimate\ldots\ smoke cigarettes, drink alcohol, get drunk at least
once a week, use marijuana, drive after having two or more drinks in
the previous hour, exceed speed limits, and do not use safety belts
(none, a few, some, most, all).'' The assessment was made at 4 time
points (baseline, 6, 12 and 18 months), the 4 scores were averaged for
each driver, and the average score was then dichotomized according to
the median split among all drivers in the study. Night driving was
determined from video data by visual observation of the ambient natural
lighting at the start of the trip. Late night was defined as 10 pm to 6
am, and night trips that did not start at late night were considered
early night trips. The presence and relative age (adult or teen) of
each passenger was determined by the coders examining the video data.
In summary, gender and risky friends enter the marginal model as
subject-level binary covariates, time of day is a~trip-level covariate
with 3 levels (day, early night and late night), passenger condition is
also a trip-level covariate with 3 levels (none, teens, at least one
adult), calendar month is a trip-level covariate with 12 levels, and
time since licensure is represented by the same set of regression
splines as before. No signs of model misspecification are observed in
residual plots (not shown).

Table \ref{tb6} presents estimates of incidence rate ratios (IRR) that
quantify the association of each $g$-force event with the aforementioned
covariates (except calendar month and time since licensure). An IRR is
just the result of exponentiating the corresponding regression
coefficient in the marginal model. The estimates are obtained using
standard GEE methods (working independence or the GOUP model, robust or
model-based variance estimate) as well as the proposed WCR-SB method
(working independence, $B=100$, $S=50$, $L=50$). All of these methods
include FSE. In a sensitivity analysis, a different separation size
($S=100$) for the WCR-SB method is used to produce similar results,
which are therefore omitted. Since the point estimates depend primarily
on the working correlation structure, only 2 sets of point estimates
are shown in Table \ref{tb6}. For the subject-level covariates, the LS
method is used to obtain 2 sets of 95\% confidence intervals (for
working independence and the GOUP model). For the trip-level
covariates, Table~\ref{tb6} presents 4 sets of 95\% confidence
intervals corresponding to the robust variance estimate and the WCR-SB
method under working independence as well as the robust and model-based
variance estimates under the GOUP model. Table \ref{tb6} contains some
unrealistically large values ($>$100, labeled as~$\infty$) for the
IRRs associating risky friends and gender with yaw and rapid start,
probably due to very few events in the subgroup of teenage drivers
acting as the denominator.

%
\begin{table}
\tabcolsep=0pt
\caption{Analysis of NTDS data using standard GEE methods (working
independence or the GOUP model, robust or model-based variance
estimate) and the proposed WCR-SB method (working independence,
$B=100$, $S=50$, $L=50$). Each outcome variable (rapid start, hard
stop, hard left/right turn, and a composite measure) is analyzed
separately in a marginal model that simultaneously adjusts for gender,
risky friends, time of day, passenger condition, calendar month1 and
time since licensure (the last two as confounders). The results are
summarized in terms of incidence rate ratios (IRRs) and 95\%
confidence intervals. The symbol $\infty$ denotes a value that is
unrealistically large ($>$100)}\label{tb6}
{\fontsize{8.5pt}{11pt}\selectfont{
\begin{tabular*}{\tablewidth}{@{\extracolsep{\fill}}lcd{2.2}ccd{2.2}cc@{}}
\hline
& &
\multicolumn{3}{c}{\textbf{GOUP model}} & \multicolumn
{3}{c@{}}{\textbf{Working independence}}\\[-4pt]
& & \multicolumn{3}{c}{\hrulefill} & \multicolumn{3}{c}{\hrulefill}\\
\multicolumn{1}{c}{\textbf{Covariate}} & \multicolumn{1}{c}{\textbf
{Comparison}}&
\multicolumn{1}{c}{\textbf{IRR}} & \multicolumn{1}{c}{\textbf
{Model-based}} & \multicolumn
{1}{c}{\textbf{Robust}} &
\multicolumn{1}{c}{\textbf{IRR}} & \multicolumn{1}{c}{\textbf{Robust}}
& \multicolumn{1}{c@{}}{\textbf{WCR-SB}}\\
\hline
\multicolumn{8}{@{}c@{}}{\textit{Composite measure} ($\mbox{\textit
{mean count}}=0.70$)}\\[4pt]
Gender&Male vs. female&1.16&\multicolumn
{2}{c}{$(0.67,1.99)$}&1.07&\multicolumn{2}{c}{$(0.64,1.79)$}\\
Risky friends&More vs. fewer&1.91&\multicolumn
{2}{c}{$(1.11,3.29)$}&1.82&\multicolumn{2}{c}{$(1.08,3.05)$}\\
Time of day&Early night vs.
day&0.64&$(0.61,0.68)$&$(0.59,0.71)$&0.73&$(0.64,0.83)$&$(0.69,0.78)$\\
&Late night vs.
day&0.76&$(0.69,0.84)$&$(0.66,0.88)$&0.89&$(0.68,1.16)$&$(0.80,1.00)$\\
Passengers&Teen vs.
none&0.81&$(0.78,0.85)$&$(0.74,0.90)$&0.78&$(0.72,0.86)$&$(0.75,0.83)$\\
&Adult vs.
none&0.39&$(0.34,0.45)$&$(0.34,0.46)$&0.32&$(0.26,0.40)$&$(0.28,0.40)$\\
[4pt]
\multicolumn{8}{@{}c@{}}{\textit{Rapid start} ($\mbox{\textit{mean
count}}=0.13$)}\\[4pt]
Gender&Male vs. female&0.12&\multicolumn
{2}{c}{$(0.00,5.69)$}&0.12&\multicolumn{2}{c}{$(0.00,5.86)$}\\
Risky friends&More vs. fewer&12.33&\multicolumn{2}{c}{$(0.26,\infty
)$}&11.50&\multicolumn{2}{c}{$(0.26,\infty)$}\\
Time of day&Early night vs.
day&0.89&$(0.83,0.95)$&$(0.70,1.14)$&0.89&$(0.74,1.07)$&$(0.83,0.95)$\\
&Late night vs.
day&0.89&$(0.79,1.00)$&$(0.68,1.15)$&0.93&$(0.77,1.12)$&$(0.78,1.03)$\\
Passengers&Teen vs.
none&0.83&$(0.78,0.88)$&$(0.68,1.01)$&0.81&$(0.70,0.92)$&$(0.75,0.87)$\\
&Adult vs.
none&0.32&$(0.25,0.41)$&$(0.21,0.50)$&0.34&$(0.26,0.44)$&$(0.26,0.44)$\\
[4pt]
\multicolumn{8}{@{}c@{}}{\textit{Hard stop} ($\mbox{\textit{mean
count}}=0.18$)}\\[4pt]
Gender&Male vs. female&0.78&\multicolumn
{2}{c}{$(0.44,1.40)$}&0.74&\multicolumn{2}{c}{$(0.42,1.30)$}\\
Risky friends&More vs. fewer&1.43&\multicolumn
{2}{c}{$(0.80,2.56)$}&1.35&\multicolumn{2}{c}{$(0.77,2.38)$}\\
Time of day&Early night vs.
day&0.68&$(0.63,0.73)$&$(0.60,0.77)$&0.70&$(0.62,0.80)$&$(0.65,0.75)$\\
&Late night vs.
day&0.71&$(0.61,0.82)$&$(0.55,0.91)$&0.70&$(0.56,0.87)$&$(0.61,0.80)$\\
Passengers&Teen vs.
none&0.90&$(0.84,0.96)$&$(0.82,1.00)$&0.88&$(0.79,0.97)$&$(0.81,0.95)$\\
&Adult vs.
none&0.51&$(0.42,0.63)$&$(0.41,0.64)$&0.38&$(0.29,0.49)$&$(0.32,0.47)$\\
\hline
\end{tabular*}
}}
\end{table}

\setcounter{table}{5}
%
\begin{table}
\tabcolsep=0pt
\caption{(Continued)}
{\fontsize{8.5pt}{11pt}\selectfont{
\begin{tabular*}{\tablewidth}{@{\extracolsep{\fill}}lccccccc@{}}
\hline
& &
\multicolumn{3}{c}{\textbf{GOUP model}} & \multicolumn
{3}{c@{}}{\textbf{Working independence}}\\[-4pt]
& & \multicolumn{3}{c}{\hrulefill} & \multicolumn{3}{c}{\hrulefill}\\
\multicolumn{1}{c}{\textbf{Covariate}} & \multicolumn{1}{c}{\textbf
{Comparison}}&
\multicolumn{1}{c}{\textbf{IRR}} & \multicolumn{1}{c}{\textbf
{Model-based}} & \multicolumn
{1}{c}{\textbf{Robust}} &
\multicolumn{1}{c}{\textbf{IRR}} & \multicolumn{1}{c}{\textbf{Robust}}
& \multicolumn{1}{c@{}}{\textbf{WCR-SB}}\\
\hline
\multicolumn{8}{@{}c@{}}{\textit{Hard left turn} ($\mbox{\textit{mean
count}}=0.20$)}\\[4pt]
Gender&Male vs. female&1.36&\multicolumn
{2}{c}{$(0.72,2.55)$}&1.29&\multicolumn{2}{c}{$(0.69,2.38)$}\\
Risky friends&More vs. fewer&2.41&\multicolumn
{2}{c}{$(1.28,4.53)$}&2.20&\multicolumn{2}{c}{$(1.19,4.07)$}\\
Time of day&Early night vs.
day&0.53&$(0.49,0.58)$&$(0.45,0.63)$&0.64&$(0.52,0.79)$&$(0.58,0.71)$\\
&Late night vs.
day&0.78&$(0.67,0.92)$&$(0.63,0.97)$&1.00&$(0.68,1.46)$&$(0.83,1.18)$\\
Passengers&Teen vs.
none&0.81&$(0.75,0.88)$&$(0.70,0.94)$&0.76&$(0.68,0.84)$&$(0.70,0.82)$\\
&Adult vs.
none&0.38&$(0.30,0.49)$&$(0.28,0.53)$&0.32&$(0.24,0.43)$&$(0.25,0.40)$\\
[4pt]
\multicolumn{8}{@{}c@{}}{\textit{Hard right turn} ($\mbox{\textit{mean
count}}=0.15$)}\\[4pt]
Gender&Male vs. female&1.39&\multicolumn
{2}{c}{$(0.72,2.69)$}&1.31&\multicolumn{2}{c}{$(0.69,2.49)$}\\
Risky friends&More vs. fewer&2.04&\multicolumn
{2}{c}{$(1.05,3.95)$}&1.92&\multicolumn{2}{c}{$(1.01,3.65)$}\\
Time of day&Early night vs.
day&0.63&$(0.58,0.67)$&$(0.50,0.79)$&0.69&$(0.55,0.85)$&$(0.62,0.76)$\\
&Late night vs.
day&0.83&$(0.74,0.93)$&$(0.46,1.48)$&0.92&$(0.58,1.46)$&$(0.75,1.14)$\\
Passengers&Teen vs.
none&0.64&$(0.60,0.69)$&$(0.54,0.76)$&0.62&$(0.55,0.70)$&$(0.56,0.69)$\\
&Adult vs.
none&0.26&$(0.20,0.33)$&$(0.18,0.37)$&0.22&$(0.16,0.31)$&$(0.17,0.30)$\\
[4pt]
\multicolumn{8}{@{}c@{}}{\textit{Yaw} ($\mbox{\textit{mean
count}}=0.04$)}\\[4pt]
Gender&Male vs. female&0.07&\multicolumn{2}{c}{$(0.00,\infty
)$}&0.07&\multicolumn{2}{c}{$(0.00,\infty)$}\\
Risky friends&More vs. fewer&\multicolumn{1}{c}{$\infty$} & \multicolumn
{2}{c}{$(0.15,\infty)$}&\multicolumn{1}{c}{$\infty$} & \multicolumn
{2}{c}{$(0.17,\infty)$}\\
Time of day&Early night vs.
day&0.92&$(0.76,1.10)$&$(0.65,1.29)$&0.94&$(0.77,1.15)$&$(0.80,1.11)$\\
&Late night vs.
day&0.88&$(0.63,1.23)$&$(0.42,1.86)$&0.92&$(0.59,1.45)$&$(0.71,1.14)$\\
Passengers&Teen vs.
none&1.19&$(1.02,1.39)$&$(0.90,1.57)$&1.17&$(1.00,1.37)$&$(1.02,1.33)$\\
&Adult vs.
none&0.54&$(0.32,0.92)$&$(0.21,1.39)$&0.44&$(0.27,0.73)$&$(0.27,0.68)$\\
\hline
\end{tabular*}
}}
\end{table}

Despite some numerical differences between the different methods, the
results in Table \ref{tb6} indicate clearly that teenage risky driving
is not associated with gender, positively associated with risky
friends, negatively associated with early night, and negatively
associated with the presence of passengers (more strongly for adults
than for teens). The evidence is not conclusive with regard to late
night, whose association with risky driving is significant in some
cases but not in others. The association of risky friends with risky
driving points to potentially destructive effects of risk-accepting
social norms and social identity on risky behavior. Both the perception
and the actuality of having risk-taking friends could contribute to the
perceived acceptability of risky behavior. The association of early
night with risky driving suggests that teens do recognize the danger of
night driving (at least partially) and drive more carefully (or rather,
less carelessly) at night. There is no contradiction between this
finding and the apparent ambiguity about late night because late night
trips often take place under unusual circumstances. Finally, it is
worth noting that passengers, especially adult passengers, appear
protective with respect to risky driving. Further research is warranted
to confirm, better characterize and utilize such protective effects.

\section{Discussion}\label{disc}

Even with a great variety of methods available for longitudinal data
analysis, it can still be difficult to analyze longitudinal data in
long sequences, especially when the serial correlation is strong and
long-lived. In this paper, we examine standard GEE methods and propose
new ones for marginal analysis of longitudinal count data in a small
number of very long sequences. The methods are evaluated and compared
in simulation experiments mimicking the NTDS, and the main findings can
be summarized as follows. We consider the use of FSE in this particular
situation, a~simple technique with important practical implications. It
allows the effects of subject-level covariates to be estimated easily
from a linear regression analysis for the estimated FSE, and it also
helps with trip-level covariates by removing the correlation due to
population heterogeneity. For trip-level covariates, we find that a
standard GEE analysis under working independence can lead to
inefficient estimates and serious undercoverage, and that both problems
can be alleviated by incorporating a properly specified correlation
structure. The latter approach works well for large counts but not for
small counts, mainly because the serial correlation is hard to estimate
with small counts. We therefore explore an alternative approach (WCR)
for the case of small counts. The original version of WCR and an
extension to simple random sampling seem unsatisfactory, however,
because of numerical instability in repeated analyses of small samples
and a bias magnification effect of WCR that results in variance
underestimation. To address these issues, we propose an WCR-SB approach
that involves separated blocks and that performs better than all of the
previously considered methods (WCR and standard GEE).

In Table \ref{tb6} the WCR-SB analyses are not dramatically different
from the other analyses. This is reassuring to the scientists, but it
also suggests that the NTDS data set is not ideal for demonstrating the
advantage of the WCR-SB approach. To put the latter point in
perspective, we note that better performance in the frequentist sense
does not imply better results in every realization. The performance of
the WCR-SB approach has been evaluated in simulation experiments which
are, in fact, designed to mimic the NTDS. Most of the uncertainty in
designing the simulation experiments is associated with the length of
the serial correlation, which is very difficult to estimate with small
counts, as shown in Table \ref{tb3}. It is certainly possible that the
serial correlation in the NTDS is not sufficiently long-lived for the
new method to make a material difference. In that case, a better
example than the NTDS might be a proposed follow-up study that involves
100 teenage drivers to be followed for at least 3 years. The larger
amount of data from the latter study will afford a better understanding
of both the nature and the length of the serial correlation.
Furthermore, as teenage drivers gain experience and perspective over
time, their driving behavior may become less haphazard and more stable,
in which case the serial correlation will gradually become longer-lived
toward the end of the (longer) study duration. Thus, in terms of the
length of the serial correlation, the follow-up study will provide a
better opportunity than the NTDS to demonstrate the advantage of the
WCR-SB approach.

The WCR-SB approach is designed for longitudinal data in a small number
of long sequences. The strength of this approach relative to standard
GEE methods is the ability to handle strong and long-lived serial
correlation when the mean count is low. The weaknesses of the WCR-SB
approach include increased computational demand (relative to standard
GEE methods) and the need for information about the length of the
serial correlation (in order to specify the separation size $S$).
Because $\gamma$ can be very difficult to estimate, one may need to
consult the subject-matter scientist or perform a sensitivity analysis
with different values of $S$, as we did in analyzing the NTDS data (see
Section \ref{analysis}).

As mentioned earlier at the end of Section \ref{wind}, we have
actually explored several existing alternatives to the robust variance
estimate in the present situation. No appreciable improvement has been
achieved using a standard bootstrap procedure (i.e., sampling subjects
with replacement) and jackknife methods [\citet{p88}, \citet
{l90}]. A~block
bootstrap procedure [\citet{k89}] appears to work well for
short-lived serial correlation but not for long-lived serial
correlation. In a simple setting that permits closed-form calculations
of the finite-sample target of the robust variance estimate (with
sample quantities replaced by their population counterparts) under
working independence, we have found that the target can be far below
the sampling variance of the point estimate observed in simulations,
suggesting that the asymptotic theory may not provide a reasonable
approximation in this situation. Given that, it seems unlikely that a
substantial improvement can be made using the available bias correction
methods [e.g., \citet{m01}]. Another possible approach to variance
estimation is window subsampling [\citet{s96}, \citet{h00},
\citet{o07}]. A
practical difficulty with this approach is specification of the window
size, which has been found to have a~large impact on the variance
estimate. \citet{h00} have suggested taking the maximum among
variance estimates based on several window sizes, but it remains
unclear how to choose the set of window sizes for the maximization.
Maximizing over a large set of window sizes can lead to a very
conservative variance estimate, as the maximum among several
underestimators need not be an underestimator itself.

Under the WCR approach, we have considered various sampling schemes
based on time. An interesting possibility, suggested by a referee, is
to sample trips using a mechanism that involves other covariates than
time and possibly the outcome. In the latter case, appropriate
adjustments will be necessary to account for the outcome-dependent
nature of the subsample. Further research is needed to explore the
potential benefits of the more sophisticated sampling schemes.

This article has been focused on valid inference in the sense of
(nearly) correct coverage. We have not discussed the issue of
generalization from a sample of subjects to the target population,
which is always important and especially so when the number of subjects
is small. Such generalizations should be easier to justify when the
within-subject variability is large relative to the between-subject
variability, which appears to be the case in the NTDS.

\begin{appendix}
\section{\texorpdfstring{IRLS estimation of $\alpha$ following a GEE analysis~with FSE}
{IRLS estimation of alpha following a GEE analysis~with FSE}}
\label{appA}

Write\vspace*{3pt} ${\underline\nu}=(\nu_1,\ldots,\nu_n)'$ and $\widehat{\underline\nu
}=(\widehat\nu_1,\ldots,\widehat\nu_n)'$, and let $\widehat{\mathbf\Sigma
}_{\widehat{\underline\nu}|{\underline\nu}}$ denote the variance
estimate from the GEE analysis. As the notation indicates, $\widehat
{\mathbf\Sigma}_{\widehat{\underline\nu}|{\underline\nu}}$ estimates
the conditional variance ${\mathbf\Sigma}_{\widehat{\underline\nu
}|{\underline\nu}}=\var(\widehat{\underline\nu}|{\underline\nu})$
because the GEE model with FSE is conditional on $\underline\nu$. The
marginal variance of $\widehat{\underline\nu}$ is given by
\[
{\mathbf\Sigma}_{\widehat{\underline\nu}}=\var(\widehat{\underline\nu
})=\var\{\epn(\widehat{\underline\nu}|{\underline\nu})\}
+\epn\{\var(\widehat{\underline\nu}|{\underline\nu})\}\approx\sigma
_b^2{\mathbf I_n}
+\epn{\mathbf\Sigma}_{\widehat{\underline\nu}|{\underline\nu}},
\]
where $\mathbf I_n$ denotes the $n\times n$ identity matrix. A natural
estimate of ${\mathbf\Sigma}_{\widehat{\underline\nu}}$ can be obtained as
%
\begin{equation}\label{a10}
\widehat{\mathbf\Sigma}_{\widehat{\underline\nu}}=\widehat\sigma
_b^2{\mathbf I_n}+\widehat{\mathbf\Sigma}_{\widehat{\underline\nu
}|{\underline\nu}}
\end{equation}
provided a reasonable estimate $\widehat\sigma_b^2$ is available. This
suggests the following IRLS algorithm. We start by obtaining an initial
estimate of $\alpha$, say, the LS estimate. Then we proceed to the
following steps:
\begin{longlist}[Step 1.]
\item[Step 1.] Calculate a moment estimate of $\sigma_b^2$, say, the
mean squared residual minus the mean diagonal element of $\widehat
{\mathbf\Sigma}_{\widehat{\underline\nu}}$;
\item[Step 2.] Substitute the estimate of $\sigma_b^2$ into \eqref{a10}
and re-estimate $\alpha$ from a~weighted least squares analysis based
on the updated $\widehat{\mathbf\Sigma}_{\widehat{\underline\nu}}$.
\end{longlist}
These steps can be iterated a number of times or until convergence.

\section{Estimation of parameters in the covariance~matrix}\label{appB}
Let $\widehat\mu_{ij}$ denote the fitted values from a preliminary GEE
analysis without FSE. Then \eqref{60} shows that a moment
estimate of the sum of $\sigma_b^2$, $\sigma_c^2$ and $\sigma_e^2$ can
be obtained as
\[
\log\Biggl[1+\frac1N\sum_{i=1}^n\sum_{j=1}^{k_i}\biggl\{
\frac{(Y_{ij}-\widehat\mu_{ij})^2}{\widehat\mu_{ij}^2}+\frac{1}{\widehat
\mu_{ij}}\biggr\}\Biggr],
\]
where\vspace*{1pt} $N=\sum_{i=1}^nk_i$ is the total number of observations. Further,
equation~\eqref{70} suggests that $\sigma_b^2$, $\sigma_c^2$ and $\gamma
$ can be estimated from a nonlinear regression analysis with
$(Y_{ij}-\widehat\mu_{ij})(Y_{ij'}-\widehat\mu_{ij'})/(\widehat\mu
_{ij}\widehat\mu_{ij'})$ as the response variable and with mean function
\[
\exp\{\sigma_b^2+\sigma_c^2\exp(-\gamma|t_{ij}-t_{ij'}|)\}-1.
\]
The above quantities could be modified using bias correction
adjustments [e.g., \citet{d00}, Section 3.2] and/or smoothing
techniques prior to the nonlinear regression analysis. However, such
modifications have not been found helpful in our experiments. Ideally,
the nonlinear regression analysis should include all pairs of trips
within subjects. However, this may be impractical for the NTDS data
because a subject with thousands of trips would contribute millions of
pairs, resulting in too many data points. As a compromise, we propose
to base the nonlinear regression analysis on two types of pairs:
consecutive trips (with $j'=j+1$) and symmetric trips (with
$j+j'=k_i+1$). The pairs of consecutive trips are informative about
short-lived serial correlation, while the pairs of symmetric trips help
characterize the overall correlation over the entire range of the gap
time $|t_{ij}-t_{ij'}|$.

Similar techniques can be used for a preliminary GEE analysis with
fixed subject effects. With $\widehat\mu_{ij|b_i}$ denoting the fitted
values with fixed subject effects, the sum of $\sigma_c^2$ and $\sigma
_e^2$ can be estimated by
\[
\log\Biggl[1+\frac1N\sum_{i=1}^n\sum_{j=1}^{k_i}\biggl\{
\frac{(Y_{ij}-\widehat\mu_{ij|b_i})^2}{\widehat\mu_{ij|b_i}^2}+\frac
{1}{\widehat\mu_{ij|b_i}}\biggr\}\Biggr],
\]
while the separate values of $\sigma_c^2$ and $\gamma$ can be estimated
from another nonlinear regression analysis with $(Y_{ij}-\widehat\mu
_{ij|b_i})(Y_{ij'}-\widehat\mu_{ij'|b_i})/(\widehat\mu_{ij|b_i}\widehat
\mu_{ij'|b_i})$ as the response variable and with mean function
\[
\exp\{\sigma_c^2\exp(-\gamma|t_{ij}-t_{ij'}|)\}-1.
\]
These are justified by \eqref{80} and \eqref{90}, respectively.

Initial values of the unknown parameters are required for fitting the
above nonlinear regression models. In our experience, a reasonable way
to obtain initial values appears to be the following linear regression
method. Consider the case with FSE, and note that equation \eqref{90}
can be rewritten as
%
\begin{equation}\label{b10}\qquad
\log\log\epn\biggl\{\frac{(Y_{ij}-\mu_{ij|b_i})(Y_{ij'}-\mu
_{ij'|b_i})}{\mu_{ij|b_i}\mu_{ij'|b_i}}+1\biggr\}
=\log\sigma_c^2-\gamma|t_{ij}-t_{ij'}|.
\end{equation}
To take advantage of this relationship, we can allocate the pairs of
trips into bins according to the value of $|t_{ij}-t_{ij'}|$ and treat
each bin as approximately homogeneous with respect to the gap time.
Within each bin, we can replace~$\mu_{ij|b_i}$ with $\widehat\mu
_{ij|b_i}$ and expectation with sample average on the left-hand side of
\eqref{b10}, replace $|t_{ij}-t_{ij'}|$ with a typical value (say, the
median) on the right-hand side, and run a linear regression analysis
with each bin as a data point. Initial estimates of $\sigma_c^2$ and
$\gamma$ can then be obtained by exponentiating the estimated intercept
and negating the estimated slope, respectively. This linear regression
approach would not work for the case without FSE, for which we could
use as initial estimates the final estimates from a GEE analysis with FSE.
\end{appendix}

\section*{Acknowledgments}

We thank the Editor, Associate Editor and two anonymous referees
for constructive comments that have greatly improved the manu\-script.


%

\printaddresses

\end{document}